%% file: iclr2027_conference.tex
\documentclass{article} 
\usepackage{iclr2027_conference,times}

\input{math_commands.tex}

\usepackage{hyperref}
\usepackage{url}
\usepackage{microtype}
\usepackage{booktabs}
\usepackage{graphicx} 
\usepackage{amsmath}
\usepackage{multirow}
\usepackage{comment}
\usepackage{xspace}
\usepackage{listings}
\usepackage{xcolor}
\usepackage{lineno}
\usepackage[dvipsnames]{xcolor}
\usepackage{booktabs}
\usepackage[table]{xcolor}
\usepackage{tikz}

\newcommand{\effipair}{\textsc{EffiPair}\xspace}
\newcommand{\rcf}{\textsc{DEF}\xspace}
\newcommand{\rcflong}{\textsc{Differential Execution Feedback (DEF)}\xspace}

\definecolor{darkblue}{rgb}{0, 0, 0.5}
\hypersetup{colorlinks=true, citecolor=darkblue, linkcolor=darkblue, urlcolor=darkblue}

\usepackage{fancyvrb}
\providecommand{\pplus}{\ensuremath{p^{+}}}
\providecommand{\pminus}{\ensuremath{p^{-}}}
\definecolor{csStore}{HTML}{F6CF8E}  
\definecolor{csIdx}{HTML}{1F5FA8}    
\definecolor{csGray}{gray}{0.42}
\definecolor{csRule}{gray}{0.6}
\definecolor{csBg}{gray}{0.965}
\definecolor{csKw}{HTML}{1F3F7F}
\definecolor{csStr}{HTML}{8B2E16}
\newcommand{\csS}[1]{{\setlength{\fboxsep}{0pt}\colorbox{csStore}{\strut#1}}}
\newcommand{\csI}[1]{{\color{csIdx}#1}}
\newcommand{\csE}[1]{{\color{csGray}#1}}

\newsavebox{\csPairBox}
\newlength{\csCW}
\lstdefinestyle{csBase}{
  basicstyle=\ttfamily\scriptsize, columns=fullflexible, keepspaces=true, tabsize=4,
  literate={'}{{\textquotesingle}}1 {`}{{\textasciigrave}}1,  
  showstringspaces=false, breaklines=true, postbreak=\mbox{\textcolor{csGray}{$\hookrightarrow$}\space},
  frame=lines, framerule=0.4pt, rulecolor=\color{csRule}, backgroundcolor=\color{csBg},
  aboveskip=6pt, belowskip=6pt, escapeinside={(*@}{@*)},
}
\lstdefinestyle{csPy}{style=csBase, language=Python, basicstyle=\ttfamily\footnotesize,
  keywordstyle=\color{csKw}\bfseries, commentstyle=\color{csGray}\itshape, stringstyle=\color{csStr}}
\lstdefinestyle{csText}{style=csBase}

\title{\effipair: Improving the Efficiency of LLM-generated Code with Differential Execution Feedback}

\author{Samira Hajizadeh\\
Department of Electrical Engineering\\
Columbia University\\
New York, NY, USA \\
\texttt{samira.hajizadeh@columbia.edu} \\
\And
Suman Jana \\
Department of Computer Science \\
Columbia University\\
New York, NY, USA \\
\texttt{suman@cs.columbia.edu}
}

\iclrfinalcopy 
\begin{document}

\maketitle
\lhead{Preprint. Under review.}

\begin{abstract}
Large language models (LLMs) can generate functionally correct programs that differ substantially in execution efficiency. Existing inference-time optimization methods typically refine each candidate using pointwise runtime or profiling feedback, which identifies how costly an implementation is or where the cost arises, but offers limited guidance on how the computation should change. We introduce \rcflong, which compares implementations that are nearby in structural program space but separated in performance, turning their execution and implementation differences into directional optimization evidence. We demonstrate \rcf in \effipair, a training-free, test-time framework that pairs structurally similar programs with different efficiencies, distills their relative execution behavior into compact feedback, and iteratively refines a candidate pool. Across EvalPerf, Mercury, and ENAMEL, using GPT-4o mini, DeepSeek-V4.1 Flash, and GPT-5 mini, \effipair achieves the highest value on each benchmark’s official efficiency metric in all nine model–benchmark settings under matched evaluation conditions. Moreover, two contrastive refinement rounds improve efficiency over the selected initial draft in every setting while preserving or improving Pass@1. These results demonstrate the effectiveness of relational execution feedback as a lightweight signal for test-time code optimization.
\end{abstract}

\section{Introduction}
Large language models (LLMs) can generate correct programs across a wide range of programming tasks, yet these programs can differ substantially in runtime, memory consumption, and computational cost. Such differences matter when generated code is repeatedly executed, incorporated into larger software systems, or deployed at scale, where seemingly small inefficiencies can accumulate into significant overhead. Improving the efficiency of LLM-generated code is therefore an increasingly important dimension of code generation alongside functional correctness. Recent work treats program efficiency as an explicit optimization objective. Benchmarks such as EffiBench~\citep{huang2025effibenchbenchmarkingefficiencyautomatically} reveal substantial efficiency gaps between model-generated and expert implementations, motivating a growing class of inference-time methods that execute candidate programs and return runtime, memory, or profiling feedback to guide refinement ~\citep{huang2025EffiLearnerenhancingefficiencygenerated,peng2024perfcodegenimprovingperformancellm,ye2025llm4effileveraginglargelanguage}. The central challenge, however, is converting this evidence into a useful implementation change. Existing inference-time refinement approaches largely provide feedback pointwise: runtime measurements reveal how costly an implementation is, and profiles identify where that cost occurs, but neither directly indicates how the computation should change. The model must therefore infer the performance-improving transformation from an isolated diagnosis, even though alternative implementations of the same computation may already provide useful comparative evidence.

Our key observation is that alternative implementations can turn execution feedback from a diagnosis into a direction for refinement. Consider two functionally equivalent programs that are structurally similar but substantially separated in execution cost. Because much of their implementation is shared, the choices on which they differ become natural candidates for explaining how the faster program avoids costs incurred by the slower one. In contrast, arbitrarily different programs may introduce too many confounding changes to yield a clear refinement signal. We therefore seek contrasts that are nearby in structural program space but distant in performance space: similar enough for their differences to be interpretable, yet separated enough in cost to provide useful optimization evidence.

\begin{figure}[t]
\begin{center}
\includegraphics[width=1.0\textwidth, trim=0 0 0 0.5cm, clip]{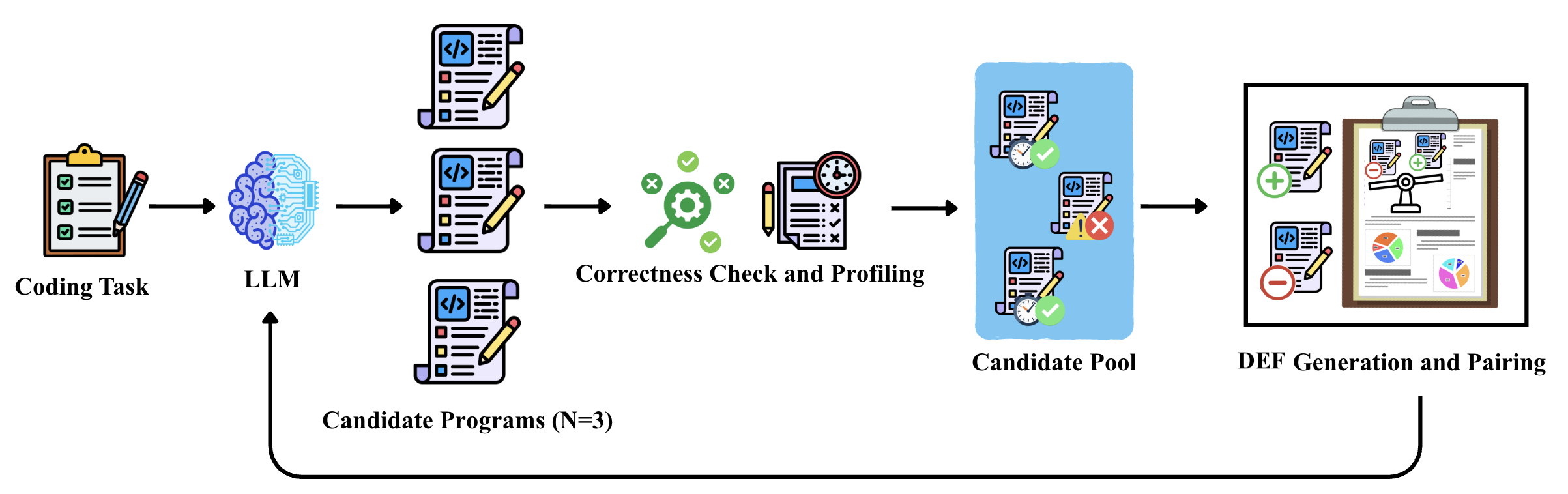}
\end{center}
\caption{\textbf{Overview of \effipair.} Given a coding task, the LLM samples $N$ candidate programs. Candidates are checked for correctness and profiled for efficiency, then stored in a candidate pool. \effipair selects a pair consisting of an efficient reference program $p^+$ and a similar but less efficient candidate $p^-$, summarizes their relative execution differences into compact \rcflong, and uses this signal to refine the program. The refined program is added back to the pool for subsequent rounds.}
\label{figure:overview}
\end{figure}

Building on this observation, we introduce \rcflong, an execution-grounded refinement signal that converts differences between an efficient program \(p^+\) and a structurally related but less efficient program \(p^-\) into directional optimization guidance. We operationalize this idea in \textbf{\effipair}, a training-free test-time framework that uses a pool of candidate programs not only to search for strong solutions, but also to construct informative contrasts for refinement. As illustrated in Figure~\ref{figure:overview}, \effipair first generates and evaluates a small candidate pool, then selects an efficient reference \(p^+\) together with a structurally similar but slower candidate \(p^-\), and constructs \rcf from the execution and implementation differences between them. The resulting signal guides the model toward performance-improving changes, after which the refined candidate is evaluated and returned to the pool for subsequent rounds. This perspective also gives candidate diversity a different role at inference time. Whereas in conventional sampling or best-of-$N$ generation, diversity is primarily useful for search, \effipair can also use an unselected candidate as optimization evidence when contrasted with a nearby, better-performing implementation. Candidate diversity thus becomes both a search resource and a source of cross-candidate refinement signals.

We further ask whether informative contrasts must be generated for the task or can instead be recovered from existing software context. At repository scale, we study contrasts from prior versions and structurally related code, testing whether \rcf extends beyond self-contained generation to optimization under real software constraints.

We evaluate \effipair across multiple code-efficiency benchmarks and strong proprietary and open-source models. Across models, \effipair improves efficiency while preserving correctness and reducing feedback overhead, and we additionally evaluate the contrastive formulation at repository scale.  Across EvalPerf, Mercury, and ENAMEL, using GPT-4o mini, DeepSeek-V4.1 Flash, and GPT-5 mini, \effipair achieves the highest value on the official efficiency metric in all 9 model–benchmark settings while preserving or improving Pass@1.

In summary, our contributions are: (1) We identify a limitation of existing inference-time execution-guided code optimization: pointwise measurements reveal \emph{how costly} an implementation is and \emph{where} that cost occurs, but provide limited guidance on \emph{how it should change}. Performance differences between functionally equivalent implementations instead provide directional optimization evidence. (2) We introduce \textbf{\rcflong}, an execution-grounded refinement signal that compares implementations nearby in structural program space but separated in performance space, turning isolated performance observations into actionable guidance. (3) We present \textbf{\effipair}, a training-free test-time framework that operationalizes \rcf through informative efficient--inefficient pairs and compact relative execution feedback. Across multiple code-efficiency benchmarks and strong language models, \effipair improves efficiency while preserving correctness and reducing feedback context. We further study repository-level optimization using generated, historical, and structurally related implementations. (4) We show that test-time candidate diversity can provide an optimization signal beyond search. Through comparisons with best-of-$N$ sampling, single-candidate profiling, paired prompting without profiling, and alternative pairing strategies, we isolate the benefit of relational execution feedback.

\section{Related Work}

Large language models (LLMs) are increasingly used for code generation, ranging from interactive prototyping to production pipelines \citep{Jiang_2026, Fakhoury_2024, jimenez2024swebenchlanguagemodelsresolve}. However, generated programs are often inefficient in runtime and memory \citep{huang2025effibenchbenchmarkingefficiencyautomatically}, leading to nontrivial cost and sustainability concerns at scale. Recent work therefore treats efficiency as an explicit optimization objective, considered alongside \citep{peng2024perfcodegenimprovingperformancellm}, or even prior to, correctness \citep{ye2025llm4effileveraginglargelanguage}. A growing set of benchmarks has been proposed to evaluate this problem \citep{waghjale2024eccoimprovemodelgeneratedcode, huang2025effibenchbenchmarkingefficiencyautomatically, du2024mercurycodeefficiencybenchmark, liu2024evaluatinglanguagemodelsefficient, qiu2025efficientllmgeneratedcoderigorous}, consistently showing that LLM-generated code lags behind expert implementations in efficiency. To address this gap, prior work explores both training-time and inference-time strategies. Training-time approaches curate performance-improving edits or incorporate efficiency-aware fine-tuning \citep{shypula2024learningperformanceimprovingcodeedits, huang2025efficoderenhancingcodegeneration}. Code-Optimise \citep{gee2025codeoptimiseselfgeneratedpreferencedata} additionally constructs correctness- and efficiency-based preference pairs for preference-based training. Inference-time methods instead rely on iterative refinement using execution feedback. 
\cite{waghjale2024eccoimprovemodelgeneratedcode} study execution-guided refinement in ECCO, which feeds correctness, runtime, and memory measurements from test executions back to the model. EffiLearner \citep{huang2025EffiLearnerenhancingefficiencygenerated} feeds runtime and memory profiles back to the model, while PerfCodeGen \citep{peng2024perfcodegenimprovingperformancellm} uses execution feedback from test runs to guide refinement. LLM4EFFI \citep{ye2025llm4effileveraginglargelanguage} further structures this process by separating algorithm selection, implementation optimization, and correctness refinement. More broadly, compiler feedback has also been incorporated into neural code generation to improve compilability \citep{wang-etal-2022-compilable}. At repository scale, recent work has explored LLM-based optimization of real-world codebases. ECO \citep{lin2025ecollmdrivenefficientcode} mines historical performance-improving commits and retrieves similar code locations to transfer optimization patterns, while PerfAgent \citep{deng2026perfagentprofilerguidediterativerefinement} iteratively guides a coding agent using summarized profiler and correctness feedback. Benchmarks such as  SWE-Perf \citep{he2026sweperflanguagemodelsoptimize} and GSO \citep{shetty2025gsochallengingsoftwareoptimization}  have been introduced to evaluate code efficiency at repository scale.

Across program- and repository-level execution-guided refinement, feedback is typically pointwise, using runtime or profiling information from the current candidate in isolation. These signals reveal cost or bottlenecks but offer limited guidance on how the implementation should change. We instead compare structurally similar candidates with different efficiency profiles to derive directional evidence for refinement. Similarity-based pairing turns candidate diversity into a cross-candidate optimization signal, while \effipair jointly preserves correctness and improves efficiency \citep{ye2025llm4effileveraginglargelanguage,peng2024perfcodegenimprovingperformancellm, huang2025EffiLearnerenhancingefficiencygenerated}.

\section{Methodology}

\effipair improves the efficiency of LLM-generated code through \rcflong. Rather than refining a program from its execution behavior in isolation, \rcf compares implementations that are structurally similar but differ substantially in execution efficiency. The central design principle is therefore to select programs that are \emph{nearby in structural program space but distant in performance space.} Structural proximity makes the contrast interpretable: when two programs share most of their implementation, the relatively small set of choices on which they differ provides localized evidence about which design decisions may account for the performance gap. At the same time, a sufficiently large difference in execution cost makes the comparison informative. In contrast, comparing arbitrarily different programs introduces many simultaneous implementation changes, making it difficult to determine which differences are relevant to efficiency. Appendix Figure~\ref{fig:case552-pair} provides an example of this process, showing a highly similar program pair with a substantial runtime gap and the resulting refinement. Using this contrast, \rcf reduces the faster program from 899.6 to 100.3 \textit{ms} in two rounds, illustrating how nearby implementations can provide actionable optimization guidance.


\subsection{Problem Setup}

Consider a programming task $\mathcal{T}(d,\mathcal{S})$, where $d$ is a natural-language task description and $\mathcal{S}=\{s_i\}_{i=1}^{|\mathcal{S}|}$ is a set of correctness tests. For a candidate program $p$, let $s_i(p)\in\{0,1\}$ indicate whether $p$ passes the $i$-th test, and let $e(p)$ denote its efficiency cost, such as execution time. Our objective is to find the most efficient functionally correct program:

$$
p^*=\arg\min_p e(p)
\quad \text{s.t.} \quad
\forall i,\; s_i(p)=1.
$$

\effipair approaches this objective through repeated propose--evaluate--refine rounds. It maintains a pool of candidate programs, evaluates them for correctness and efficiency, constructs informative efficient--inefficient pairs, and derives \rcf from their relative execution behavior to guide subsequent refinement.

\subsection{Candidate Generation and Pair Selection}
\label{method:generationandpairingstrategy}

We implement \effipair as an iterative propose--evaluate--refine pipeline. Given a task description $d$, we first generate an initial pool of candidates,

$$
C_0=\{p_0^{(j)}\}_{j=1}^{N},
$$

by independently prompting the LLM with $d$ without any additional guiding signals and requesting a code solution. Sampling the model $N$ times produces diverse candidate implementations that can subsequently serve both as potential solutions and as contrasts for refinement. At round $t$, each candidate in the current pool is evaluated for functional correctness and execution efficiency. For each candidate $p$, we additionally construct an abstract syntax tree (AST) representation $A(p)$, which is used to measure structural similarity between programs. We first select the most efficient correct candidate as the reference program $p^+$:

$$
p^+
=
\arg\min_{p\in C_t} e(p)
\quad
\text{s.t.}
\quad
\forall i,\; s_i(p)=1.
$$

We then construct a less efficient but structurally similar partner $p^-$. Let

$$
\mathrm{sim}_{\mathrm{AST}}(p,q)
=
\cos\!\big(A(p),A(q)\big)
$$

denote the AST similarity between two programs. Given a similarity threshold $\tau$, we define the set of sufficiently similar correct candidates

$$
\mathcal{N}(p^+)
=
\left\{
q\in C_t :
q\neq p^+,\;
\forall i,\; s_i(q)=1,\;
\mathrm{sim}_{\mathrm{AST}}(p^+,q)\geq\tau
\right\}.
$$

Among these candidates, we select the least efficient program:

$$
p^-
=
\arg\max_{q\in\mathcal{N}(p^+)} e(q).
$$

This selection produces a pair that is structurally close but separated in execution efficiency: the similarity constraint limits unrelated implementation differences, while choosing the slowest candidate within the neighborhood provides a sufficiently strong performance contrast for constructing DEF. The formulation applies when at least two functionally correct candidates are available. As a correctness-recovery fallback, when only one correct candidate is available and a standard correct–correct DEF pair cannot be constructed, we retain the correct candidate as Candidate A and select a structurally similar incorrect candidate as Candidate B, along with its failure information. In this setting, the resulting feedback is used primarily to preserve correctness while searching for an improved implementation, rather than to infer an efficiency contrast between the two candidates. If no correct candidate is available, two incorrect candidates and their failure information are provided to the model, which is instructed to restore correctness before optimizing efficiency.

\subsection{Iterative Refinement}
Given the selected pair $(p^+,p^-)$, \effipair constructs \rcf from their implementation and execution differences. Rather than providing performance information for either program in isolation, \rcf provides the model with a contrastive signal between the two candidates, relating the implementation choices on which they differ to their observed difference in execution efficiency. This provides the LLM with a directional signal about which parts of the less efficient implementation can be changed while preserving the behavior of the efficient reference. Let $\Delta(p^+,p^-)$ denote the resulting DEF signal. Given the task description $d$, the selected pair $(p^+,p^-)$, and their differential feedback, the refinement step produces a new candidate

$$
\tilde{p}
=
R\bigl(d,p^+,p^-,\Delta(p^+,p^-)\bigr),
$$

where $R$ denotes the LLM-based refinement procedure. The model is instructed to preserve functional correctness while using the differences between the two implementations to identify performance-improving changes. The refined candidate $\tilde{p}$ is then evaluated for correctness and efficiency and added back to the candidate pool. Pair selection and refinement are repeated on the updated pool in subsequent rounds, allowing newly generated candidates to become either stronger reference programs or informative contrast candidates. Because previously generated candidates are retained, an unsuccessful refinement does not discard the best correct solution already found.

\section{Experiments}

\subsection{Setup}
We evaluate \effipair on a suite of code-generation benchmarks designed to assess not only functional correctness but also the execution efficiency of generated programs. Our analysis emphasizes benchmark-specific metrics which represent relative improvement, rather than absolute execution times, since raw latency is sensitive to hardware, operating system scheduling, and library versions. Specifically, we study the effect of \effipair on three code-efficiency benchmarks: Mercury \citep{du2024mercurycodeefficiencybenchmark}, ENAMEL \citep{qiu2025efficientllmgeneratedcoderigorous}, and EvalPerf \citep{liu2024evaluatinglanguagemodelsefficient}. EvalPerf targets performance-challenging tasks and reports \textbf{DPS}, which measures a solution’s efficiency relative to the nearest slower reference solution, with the normalized version \textbf{$\text{DPS}_{\mathrm{norm}}$}. Mercury uses the \textbf{Beyond} metric, which normalizes runtime percentiles over each task’s runtime distribution to enable hardware-robust comparisons. ENAMEL reports \textbf{eff@1}, a weighted score based on the worst execution time across test cases of different difficulty levels; higher values indicate better efficiency, and values above 1 mean the generated code outperforms the expert solution. We also report \textbf{Pass@1} for all runs. We use a diverse set of language models to examine whether \effipair can improve systems that already have strong efficiency and correctness performance. Specifically, our evaluation includes GPT-5.6 Luna, GPT-5 mini, GPT-4o mini \citep{openai2026gpt56, openai2024gpt4omini, openai2025gpt5}, Claude Sonnet 5 \citep{anthropic2026sonnet5}, DeepSeek-V4.1 Flash \citep{deepseek2026v41flash}, and Gemini 3.8 Flash \citep{google2026gemini38}. This setting provides a stringent test of whether \effipair remains effective even when the underlying model is already highly capable.

\begin{table}[t]
    \centering
    \small
    \setlength{\tabcolsep}{3pt}
    \begin{tabular}{ll|rrrrrrrrr}
    \toprule
    & & \multicolumn{3}{c}{EvalPerf} & \multicolumn{3}{c}{Mercury} & \multicolumn{3}{c}{ENAMEL} \\
    \cmidrule(lr){3-5}\cmidrule(lr){6-8}\cmidrule(lr){9-11}
    Model & Method & DPS$_{\mathrm{norm}}$ & Pass@1 & Tok. & Beyond@1 & Pass@1 & Tok. & eff@1 & Pass@1 & Tok. \\
    \midrule
    \multirow{10}{*}{\shortstack[l]{GPT-4o\\ mini}} & Baseline (instruct) & 75.50 & 89.83 & 0.2 & 81.06 & 92.19 & 0.5 & 49.78 & 83.80 & 0.3 \\
     & EffiLearner-Gen & 80.28 & \textbf{88.98} & 0.6 & 79.94 & 90.62 & 0.5 & 48.51 & 83.10 & 0.6 \\
     & ECCO nl2code & 79.49 & 82.20 & 0.3 & 79.68 & 89.45 & 0.5 & 53.99 & 83.80 & 0.7 \\
     & ECCO exec-refine & 74.88 & 67.80 & 0.7 & 80.32 & 89.84 & 1.2 & 47.99 & 75.35 & 1.3 \\
     & ECCO self-refine & 74.59 & 61.02 & 2.1 & 74.49 & 85.16 & 2.6 & 36.22 & 60.56 & 2.9 \\
     & ECCO nl+exec-refine & 73.49 & 50.85 & 2.0 & 76.22 & 86.33 & 2.7 & 38.76 & 68.31 & 2.9 \\
     & EffiLearner-Opt & 74.34 & 86.44 & 6.3 & 82.84 & 91.80 & 9.7 & 54.80 & 83.80 & 5.6 \\
     & LLM4EFFI & 81.08 & 72.88 & 59.1 & 76.85 & 85.16 & 79.6 & 42.52 & 71.13 & 78.3 \\
     & \effipair $r_0$ & 83.58 & 88.14 & 1.4 & 86.32 & \textbf{95.18} & 1.7 & 57.50 & \textbf{86.62} & 1.9 \\
     & \effipair $r_2$ & \textbf{84.25} & \textbf{88.98} & 3.0 & \textbf{86.62} & \textbf{95.18} & 3.9 & \textbf{59.71} & \textbf{86.62} & 5.1 \\
    \midrule
    \multirow{10}{*}{\shortstack[l]{DeepSeek-\\ V4.1 Flash}} & Baseline (instruct) & 81.48 & 88.14 & 0.2 & 89.17 & \textbf{98.05} & 0.5 & 56.38 & 93.66 & 0.3 \\
     & EffiLearner-Gen & 81.88 & 88.14 & 0.4 & 88.98 & 97.27 & 0.5 & 58.70 & \textbf{94.37} & 0.7 \\
     & ECCO nl2code & 84.36 & 82.20 & 0.3 & 89.51 & 95.70 & 0.5 & 68.45 & 90.85 & 0.9 \\
     & ECCO exec-refine & 82.97 & 83.90 & 0.7 & 89.81 & 95.70 & 1.0 & 71.04 & 92.96 & 1.4 \\
     & ECCO self-refine & 80.05 & 80.51 & 1.9 & 85.46 & 93.75 & 3.1 & 63.41 & 85.92 & 3.4 \\
     & ECCO nl+exec-refine & 80.10 & 75.42 & 2.1 & 87.07 & 94.14 & 3.0 & 66.85 & 86.62 & 3.4 \\
     & EffiLearner-Opt & 85.79 & 85.59 & 4.3 & 92.44 & 96.88 & 8.7 & 73.06 & \textbf{94.37} & 4.5 \\
     & LLM4EFFI & 85.36 & 89.83 & 35.7 & 83.09 & 88.28 & 42.5 & 75.86 & 88.73 & 44.8 \\
     & \effipair $r_0$ & 87.59 &  88.98 & 1.1 & 92.57 & 97.79 & 1.5 & 71.16 & \textbf{94.37} & 2.2 \\
     & \effipair $r_2$ &  \textbf{89.50} & \textbf{91.53} & 2.6 & \textbf{94.24} & 97.79 & 3.7 & \textbf{77.58} & \textbf{94.37} & 5.2 \\
    \midrule
    \multirow{10}{*}{\shortstack[l]{GPT-5\\ mini}} & Baseline (instruct) & 79.56 & 86.44 & 0.2 & 89.78 & 96.48 & 0.6 & 56.19 & 94.37 & 0.4 \\
     & EffiLearner-Gen & 82.34 & 86.44 & 0.4 & 88.31 & 95.70 & 0.6 & 57.82 & \textbf{95.77} & 0.4 \\
     & ECCO nl2code & 82.35 & 85.59 & 0.4 & 88.32 & 95.31 & 0.6 & 61.48 & 92.25 & 0.6 \\
     & ECCO exec-refine & 76.48 & 83.05 & 1.1 & 89.12 & 95.70 & 1.3 & 62.60 & 91.55 & 1.2 \\
     & ECCO self-refine & 74.84 & 70.34 & 2.9 & 87.76 & 95.70 & 3.3 & 45.70 & 73.24 & 3.4 \\
     & ECCO nl+exec-refine & 78.53 & 67.80 & 3.1 & 87.70 & 94.92 & 3.2 & 55.58 & 84.51 & 3.1 \\
     & EffiLearner-Opt & 82.92 & 87.29 & 5.7 & 91.40 & 95.70 & 9.4 & 70.35 & 94.37 & 4.8 \\
     & LLM4EFFI & 73.62 & 83.05 & 36.1 & 86.38 & 92.97 & 47.4 & 63.65 & 88.03 & 43.2 \\
     & \effipair $r_0$ & 85.42 &  \textbf{89.83} & 1.0 & 91.33 & \textbf{97.27} & 1.8 & 69.74 & 95.07 & 1.2 \\
     
     & \effipair $r_2$ &  \textbf{85.95} &  \textbf{89.83} & 3.0 & \textbf{92.32} & \textbf{97.27} & 4.4 & \textbf{72.47} & 95.31 & 6.0 \\
    \bottomrule
    \end{tabular}
    \caption{\effipair versus prior efficiency-oriented methods, all reproduced with the same models, decoding settings and evaluation harnesses. Each benchmark reports its official efficiency metric and Pass@1 (\%). $r_0$ is \effipair's selected draft before refinement and $r_2$ its output after two contrastive rounds; both are means over three independent runs, while other rows are single runs. Bold marks the best efficiency and Pass@1 within each model block. Tok.\ is thousands of tokens per task (prompt $+$ completion) for that run; for \effipair it covers the three sampled drafts and, in the $r_2$ row, the two refinement rounds.}
    \label{tab:prior-methods-table}
\end{table}

\effipair uses \(N=3\) initial candidates and \(T=2\) refinement rounds by default, which provides a practical balance between candidate diversity and inference cost. The three candidates are independent samples generated from the same prompt using different seeds. The similarity window is chosen by default to be \([0.85, 1.0]\) (see Appendix \ref{appendix:similarity} for further analysis) and byte-identical partners are excluded, as they carry no contrastive signal. The similarity metric, described in Section \ref{method:generationandpairingstrategy}, is computed on abstract syntax trees. To reduce test-case overfitting, benchmark tests are not included in the prompt for our method (\effipair); instead, the model receives only compact execution and profiling feedback from \texttt{cProfile} \citep{python_cprofile}, which is a built-in Python profiler. We compact the profiler output into a short summary; implementation details are provided in Appendix~\ref{appendix:profiling-detail-compact}, and the rationale for condensed profiling is discussed in Appendix~\ref{appendix:whycondensedprofiling}. To avoid selection bias from runtime noise, candidate selection and final evaluation use separate measurements. After \effipair selects the fastest correct candidate using its internal profiling measurements, the selected program is independently re-executed under the benchmark’s official evaluation harness, and only this independent measurement is used for reported efficiency results. On EvalPerf, \effipair's internal pool gate uses only the EvalPlus base suite (a mean of $3.9$ inputs per task), while final scoring uses the official harness over base $+$ plus ($285$ inputs per task); EffiLearner's optimization stage gates on the same base suite, and ECCO's execution-refining variants use at most three cases. On Mercury, \effipair validates against the benchmark's shipped \texttt{test\_cases} through the same builder the official scorer uses, as do ECCO and EffiLearner. On ENAMEL, \effipair and ECCO's execution-refining variants share an identical gate---$16$ of the scorer's $20$ cases under the official seeds, while EffiLearner draws the same number of cases under different seeds. LLM4EFFI touches no benchmark tests on any of the three and uses $20$ self-made test cases per task. All experiments were run on dedicated AWS EC2 \texttt{c7i.2xlarge} instances (Intel Xeon Scalable ``Sapphire Rapids'', 3.2\,GHz sustained; 4 physical cores / 8 logical threads, 16\,GiB RAM, 80\,GB gp3 EBS) in the \texttt{us-east-1} region, running Ubuntu 24.04 LTS (kernel \texttt{7.0.0-1012-aws}) with Python 3.12.3.

\subsection{Baselines and Comparison Methods}

We compare \effipair against three published efficiency-oriented systems, each reproduced from its official implementation under the same models, decoding settings and evaluation harnesses. \textbf{ECCO}~\citep{waghjale2024eccoimprovemodelgeneratedcode} refines a single generated program from feedback; we include all three of its refinement modes -- natural-language self-critique, execution feedback, and natural-language reflection over execution feedback -- alongside its direct \textit{nl2code} generation. \textbf{EffiLearner}~\citep{huang2025EffiLearnerenhancingefficiencygenerated} profiles the current program with line- and memory-level instrumentation, returns the overhead report to the model, and accepts a candidate only if it executes and lowers the measured memory integral; we report both its initial generation and its output after five self-optimization iterations. \textbf{LLM4EFFI}~\citep{ye2025llm4effileveraginglargelanguage} instead searches: it formalizes the task, proposes several algorithms and implementations, synthesizes its own test cases to verify them, and selects a final program by model judgment. As a lower bound we include an instruction-only baseline at temperature~0, using LLM4EFFI's own direct-generation prompt and sanitizer. Together these span the design space around \effipair (single-call generation, feedback-driven refinement of one program, and search over many candidates) and they differ by more than two orders of magnitude in inference cost, which is why we report tokens per task alongside every efficiency number.

\begin{table}[t]
\centering
\small
\setlength{\tabcolsep}{3.5pt}

\begin{tabular}{l|cccc}
\toprule
\textbf{Method}
& \shortstack{\textbf{DeepSeek-V4.1}\\\textbf{Flash}}
& \shortstack{\textbf{Gemini 3.8}\\\textbf{Flash}}
& \shortstack{\textbf{GPT-5.6}\\\textbf{Luna}}
& \shortstack{\textbf{Claude}\\\textbf{Sonnet 5}} \\
& \scriptsize eff@1 / Pass@1
& \scriptsize eff@1 / Pass@1
& \scriptsize eff@1 / Pass@1
& \scriptsize eff@1 / Pass@1 \\
\midrule

\multicolumn{5}{l}{\textit{Generation}} \\

\quad Greedy Draft
& 55.73 / 93.66
& 60.88 / 97.18
& 55.29 / 90.14
& 56.01 / 94.37 \\

\quad Single Sampled Draft
& 65.30\,${\scriptstyle \pm 2.34}$ / 93.66
& 78.99\,${\scriptstyle \pm 1.34}$ / 96.71
& 64.36\,${\scriptstyle \pm 0.71}$ / 89.20
& 62.76\,${\scriptstyle \pm 0.96}$ / 93.66 \\

\quad Best-of-5
& 72.61\,${\scriptstyle \pm 1.99}$ / 93.19
& 85.23\,${\scriptstyle \pm 0.96}$ / \textbf{98.12}
& 71.58\,${\scriptstyle \pm 1.23}$ / \textbf{94.37}
& 69.09\,${\scriptstyle \pm 1.74}$ / \textbf{96.24} \\

\quad Pool Selection ($r_0$)
& 71.52\,${\scriptstyle \pm 1.32}$ / 94.84
& 83.70\,${\scriptstyle \pm 0.81}$ / 97.42
& 69.91\,${\scriptstyle \pm 1.06}$ / 93.66
& 67.38\,${\scriptstyle \pm 1.20}$ / 94.60 \\

\addlinespace[4pt]
\multicolumn{5}{l}{\textit{+ refinement}} \\

\quad Isolated Refine
& 75.11\,${\scriptstyle \pm 1.52}$ / 95.54
& 84.81\,${\scriptstyle \pm 0.36}$ / 97.42
& 70.75\,${\scriptstyle \pm 0.53}$ / 93.90
& 70.64\,${\scriptstyle \pm 0.91}$ / 95.07 \\

\quad Random-Pair Refine
& 76.25\,${\scriptstyle \pm 1.80}$ / 96.24
& 87.65\,${\scriptstyle \pm 0.72}$ / 97.18
& 71.47\,${\scriptstyle \pm 1.48}$ / 93.66
&  69.09\,${\scriptstyle \pm 1.30}$ / 94.37 \\

\quad Matched-Pair Refine
& 77.95\,${\scriptstyle \pm 1.54}$ / 96.01
& 88.64\,${\scriptstyle \pm 0.28}$ / 97.42
& 71.70\,${\scriptstyle \pm 1.10}$ / 93.66
& 70.39\,${\scriptstyle \pm 0.58}$ / 94.60 \\

\addlinespace[4pt]
\multicolumn{5}{l}{\textit{+ execution feedback}} \\

\quad Isolated Refine + EF
& 79.09\,${\scriptstyle \pm 1.08}$ / 96.60
& 90.05\,${\scriptstyle \pm 0.40}$ / 97.18
& 75.18\,${\scriptstyle \pm 0.65}$ / 93.90
&  72.79\,${\scriptstyle \pm 2.10}$ / 94.37 \\

\quad Random-Pair Refine + EF
& 79.92\,${\scriptstyle \pm 1.61}$ / 96.01
& 90.18\,${\scriptstyle \pm 0.18}$ / 97.42
& 74.07\,${\scriptstyle \pm 1.41}$ / 93.90
& 73.86\,${\scriptstyle \pm 1.38}$ / 94.37 \\

\quad \effipair
& \textbf{80.25\,${\scriptstyle \pm 1.27}$} / \textbf{96.71}
& \textbf{90.93\,${\scriptstyle \pm 0.51}$} / 97.42
& \textbf{75.34\,${\scriptstyle \pm 0.47}$} / 93.90
& \textbf{73.93\,${\scriptstyle \pm 1.23}$} / 94.60 \\

\bottomrule
\end{tabular}

\caption{
Component ablation on ENAMEL (142 tasks), reporting the official
eff@1 / Pass@1 (\%) metrics across four model families.
We compare generation-only baselines with isolated refinement, random-pair refinement, and matched-pair refinement, with and without the corresponding constructed execution feedback (EF). In the last row, \rcf is implemented by combining matched-pair refinement with execution feedback to create \effipair. Best-of-5 is included as a generation-only control. All refinement methods operate on the same initial candidate pool and use the same correctness gate. Bold denotes the best reported value for each metric within each model.}

\label{tab:ablation-enamel}
\end{table}

\subsection{Main results}

\begin{figure}[t]
\begin{center}
\includegraphics[width=1\textwidth, trim=0 0 0 0.4cm, clip]{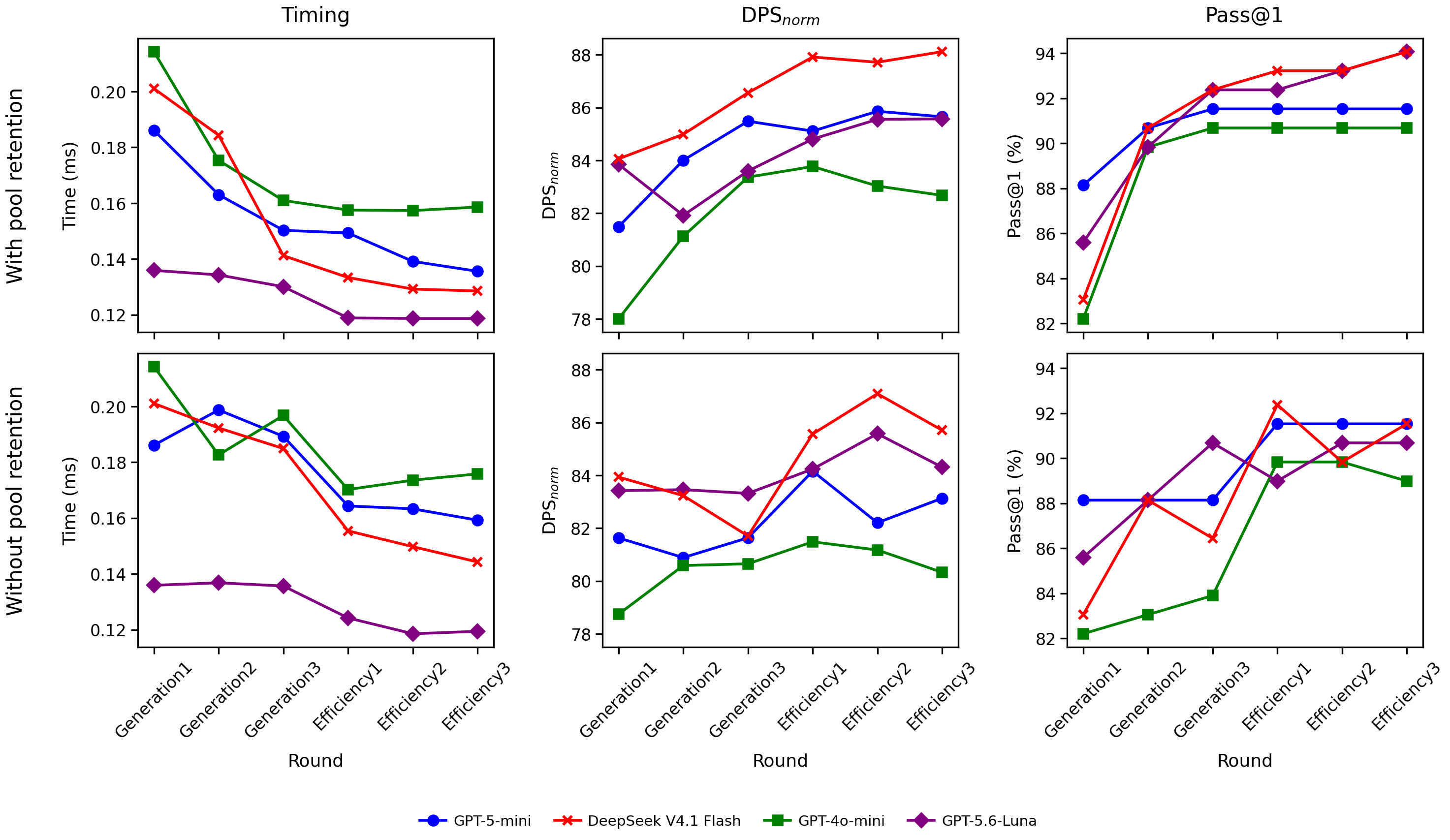}
\caption{Iterative refinement behavior of \effipair on EvalPerf across evaluated models.}
\label{figure:iterative-refinement}
\end{center}
\end{figure}

Table \ref{tab:prior-methods-table} establishes the end-to-end effectiveness of \effipair. \effipair achieves the highest official efficiency across nine settings after two refinement rounds ($r_2$), indicating that the benefit is not tied to a particular model family or efficiency metric. The comparison between $r_0$ and $r_2$ is particularly important since it isolates the contribution of \rcf beyond the gains from initial candidate selection. Across all settings $r_2$ improves over $r_0$, which further reinforces \effipair's central claim that the refinement stage contributes beyond candidate selection: sampling three candidates and keeping the fastest correct one does not exhaust the available efficiency, and the subsequent rounds recover gains that selection alone leaves behind. Subsection ~\ref{subsec:ablation} further isolates the impact of each component. Importantly, Table \ref{tab:prior-methods-table} shows that the efficiency gains do not come at the expense of functional correctness. Across the nine model--benchmark settings in Table~\ref{tab:prior-methods-table}, Pass@1 never decreases from $r_0$ to $r_2$: it is preserved in six settings and improves in three. Figure~\ref{figure:iterative-refinement} provides further evidence for this behavior across refinement rounds: Pass@1 remains stable or improves as efficiency optimization proceeds. Together, these results show that \effipair can continue extracting efficiency gains while preserving the correctness established during candidate generation and selection.

\subsection{Ablation Results}
\label{subsec:ablation}
Table \ref{tab:ablation-enamel} disentangles the contributions of candidate selection, pairing, and execution feedback to \effipair. With the initial pool and correctness gate held fixed, \effipair improves over \emph{Pool Selection $(r0)$} by 5.43–8.73 eff@1 points across all four model families. It also outperforms \emph{Best-of-5} in every case, showing that its gains cannot be explained by additional sampling alone. The ablations further show that both execution feedback and pair construction matter.

Adding execution feedback improves \emph{Isolated Refine} for every model, while \effipair consistently exceeds \emph{Random-Pair Refine + EF}. Relative to \emph{Matched-Pair Refine} without execution feedback, the full method gains 2.29–3.64 eff@1 points. This suggests a complementary effect: execution feedback supplies the performance signal, while similarity-based pairing focuses that signal on localized implementation differences that are more useful for refinement. These efficiency gains are achieved without sacrificing correctness: \effipair matches or improves the $r0$ Pass@1 for all four models. Overall, Table 2 shows that \effipair benefits not merely from selecting strong candidates, but from turning relationships among candidates into actionable, execution-grounded refinement signals.

\subsection{Iterative Refinement}
Figure \ref{figure:iterative-refinement} examines the behavior of \effipair over successive generation and efficiency-refinement checkpoints on EvalPerf. The top row reports the standard setting with pool retention. Across all four models, execution time decreases substantially, $DPS_{\mathrm{norm}}$ improves over the initial generation, and Pass@1 is preserved or increased. Importantly, these gains extend beyond initial candidate selection, with later efficiency rounds continuing to improve runtime and, in most cases, $DPS_{\mathrm{norm}}$. The magnitude and timing of improvement vary across models—some improve sharply in early rounds and then stabilize, while others continue to benefit from later refinement—but the overall pattern shows that \effipair extracts additional efficiency from an already strong candidate pool without sacrificing correctness. The bottom row isolates the role of pool retention by repeating the same analysis without retaining previously discovered candidates. The resulting trajectories are noticeably less stable. A refinement that improves a metric at one checkpoint can be followed by a worse rewrite at the next: execution time increases in several transitions, $DPS_{\mathrm{norm}}$ can fall after reaching a stronger value, and Pass@1 similarly exhibits regressions after earlier gains. In the standard \effipair loop, refined candidates are instead added back to the existing pool, so a successful candidate remains available even when a later refinement is unsuccessful. Pool retention therefore turns iterative refinement into a best-so-far process: once \effipair discovers a strong correct solution, subsequent rounds can improve upon it without requiring that solution to be discarded. The contrast between the two rows shows that retention is important not only for obtaining good candidates, but also for preventing later optimization steps from undoing gains already found.

\subsection{Repository-level experiments}
To test \effipair in a more realistic setting, we evaluate it at repository scale, where a candidate is a patch against a live codebase rather than a self-contained function. We use \textbf{SWE-Perf}, a benchmark of 140 instances derived from real performance-improving pull requests across nine widely used Python repositories. Each instance provides a repository snapshot at the pre-merge commit, a set of performance tests, the expert patch that was actually merged, and---in the \emph{oracle} setting we adopt---the names of the functions that patch modified. We score every arm with SWE-Perf's own unmodified evaluation harness, which times each performance test over repeated runs in a fresh container and reports a gain only when a one-sided Mann--Whitney test distinguishes it from run-to-run noise. This setting changes what \effipair must do. At function level, the unit of generation, the unit of measurement, and the unit of edit coincide; in a repository they come apart, and a candidate must preserve the surrounding API and pass the project's own tests before its runtime is even measurable. It also changes where contrastive pairs can come from: rather than resampling the same function, a repository supplies its own history and its own structurally similar code. Our experiments hold the model, the instances, and the meter fixed and vary only the source of the contrast, isolating the contribution of \emph{pair quality} from that of refinement itself.

\begin{table}[t]
\centering
\small
\setlength{\tabcolsep}{5pt}
\begin{tabular}{llrr}
\toprule
Arm & Contrast source & $n$ & Mean gain (\%) \\
\midrule
Best seed (no refinement) & none & 125 & 1.21 \\
\addlinespace[2pt]
Solo (profile only) & none & 125 & 1.19 \\
Same-site pair & resampled variant & 125 & 1.24 \\
History pair & own prior version & 107 & 1.16 \\
Sibling pair & similar function & 79 & 0.85 \\
History $+$ sibling & both & 119 & 1.06 \\
Retrieved exemplar & BM25 perf-commit & 125 & 1.17 \\
\addlinespace[2pt]
\quad \emph{ablation:} no pool retention & none & 120 & 0.76 \\
\addlinespace[2pt]
\bottomrule
\end{tabular}

\caption{\effipair at repository scale on SWE-Perf (125 of 140 instances; exclusions in Appendix~\ref{app:swe-perf-exclusions}). Each arm commits one patch per instance, the fastest correct member of its pool on \effipair's own internal clock, which is not the metric reported here, and all arms and the expert patch are then measured together in one scoring session; \textbf{Mean gain} is the significance-gated improvement under SWE-Perf's unmodified harness over that row's $n$ instances.}
\label{tab:repo-level}
\end{table}

\section{Limitations}
\effipair has several limitations. Its effectiveness depends on identifying informative candidate pairs: limited diversity, weak structural similarity, or noisy execution measurements can reduce the quality of the refinement signal. In addition, \effipair currently optimizes primarily for correctness and execution efficiency, without explicitly considering software qualities such as readability and maintainability. In some cases, performance-oriented refinements may produce more specialized or compact implementations that are harder to understand or modify, creating a trade-off between efficiency and code quality. \effipair also introduces additional generation, execution, and profiling overhead, making it most suitable when the optimized code is executed frequently enough to justify this cost. Future work could incorporate readability, maintainability, and other software-quality objectives into the refinement process, while also improving pair selection and adaptive similarity criteria.

\section{Conclusion}
We introduced \effipair, a training-free test-time framework that uses \rcflong to improve the efficiency of LLM-generated code. By contrasting structurally similar programs with different execution behavior, \effipair turns candidate diversity into a useful signal for refinement. Across multiple benchmarks and language models, this approach consistently improves efficiency while preserving functional correctness. Overall, our results suggest that DEF provides a practical and general mechanism for guiding test-time code optimization.

\subsection*{AI use statement}

Generative AI tools were used as writing assistants to improve the clarity, grammar, phrasing, and overall readability of author-written text. All AI-assisted text was reviewed and edited by the authors, who take responsibility for the final content of the paper.



\bibliography{iclr2027_conference}
\bibliographystyle{iclr2027_conference}

\appendix
\section{Appendix}





\begin{figure}[t]
\begin{center}
\includegraphics[width=0.7\textwidth]{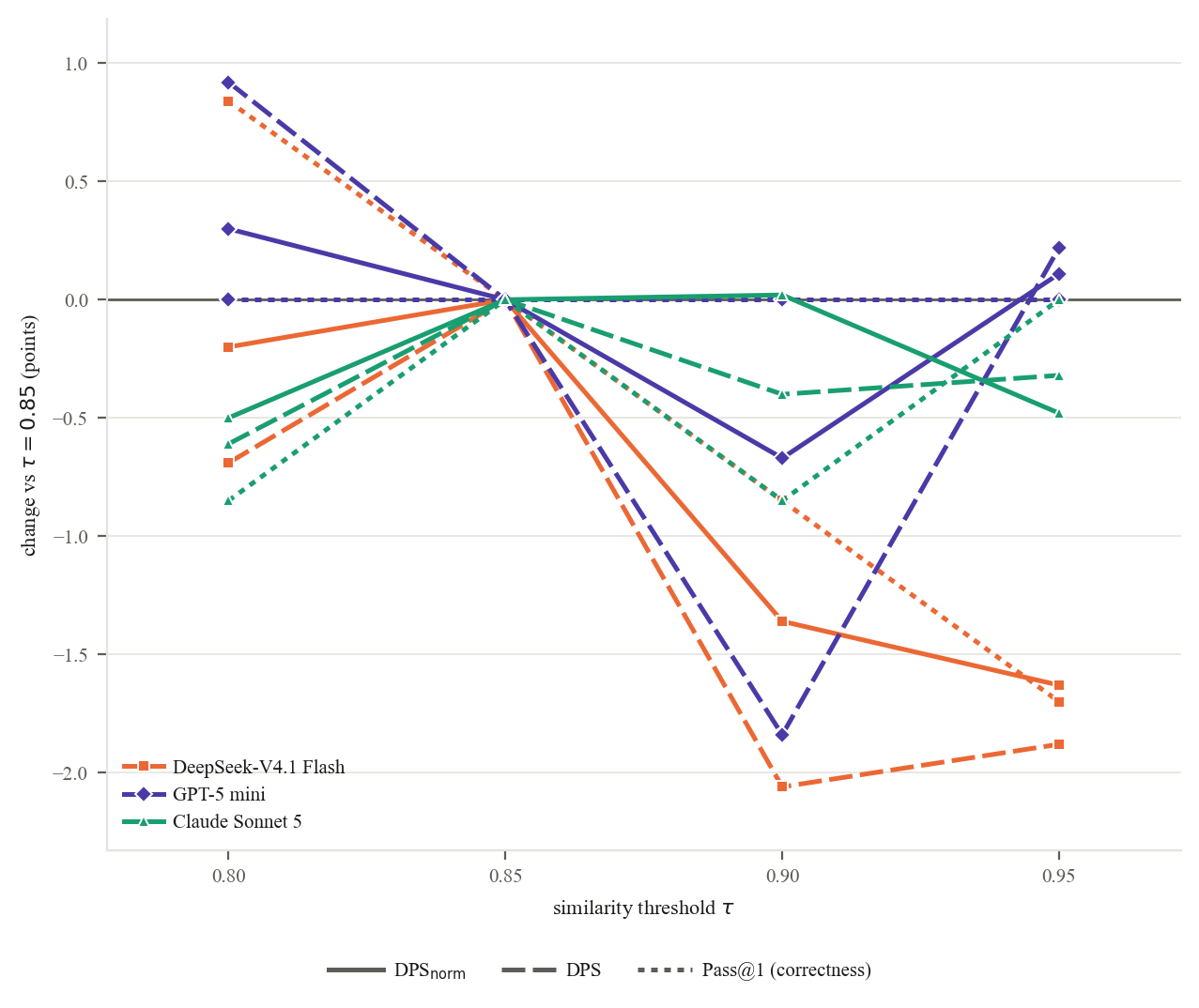}
\caption{DPS, $\text{DPS}_\text{norm}$, and Pass@1 for various similarity thresholds on the EvalPerf benchmark.}
\label{figure:tau}
\end{center}
\end{figure}

\subsection{Similarity}
\label{appendix:similarity}

\subsubsection{Similarity Threshold}
\label{appendix:TauAnalysis}

We use AST representations to measure the structural similarity between programs: each candidate is parsed with Python's \texttt{ast} module, after
stripping comments and top-level assertions, and represented as a
$132$-dimensional bag-of-node-types vector counting occurrences of every AST
node class. These vectors are $L_2$-normalized and compared by cosine
similarity, and the pairing window $[\tau, 1]$ is applied to that score. This representation is simple, inexpensive, and sufficient for capturing broad structural overlap without introducing heavy preprocessing. Figure \ref{figure:tau} shows \effipair's performance with various $\tau$ values in different settings. Our default value, $\tau=0.85$, seems to present a local optimum in terms of the balance between accuracy and efficiency across settings. 

\begin{figure}[t]
\centering
\settowidth{\csCW}{\scriptsize\ttfamily M}%
\begin{lrbox}{\csPairBox}%
\begin{minipage}{\dimexpr 109\csCW+6pt\relax}
\centering
\begin{minipage}[t]{\dimexpr 50\csCW+1pt\relax}
{\scriptsize \pplus{} (faster)\hfill\textbf{899.6\,ms}}\par\vspace{1pt}
\begin{Verbatim}[commandchars=\\\{\},fontsize=\scriptsize,frame=lines,framerule=0.4pt,rulecolor=\color{csRule},framesep=2pt,
  listparameters={\setlength{\topsep}{0pt}\setlength{\partopsep}{0pt}\setlength{\parskip}{0pt}}]
MOD = 10**9 + 7
\csS{dp = [[0]*3 for _ in range(2)]}
dp[0][0] = 1
for \csI{_} in range(\csI{n}):
    \csS{ndp = [[0]*3 for _ in range(2)]}
    for a in range(2):
        for l in range(3):
            val = dp[a][l]
            \csE{...}
    \csS{dp = ndp}
\csE{...}
\end{Verbatim}
\par\vspace{4pt}{\centering
\begin{tikzpicture}[x=\csCW,y=\csCW,every node/.style={font=\scriptsize\ttfamily,inner sep=0.6pt,text height=1.6ex,text depth=0.3ex,fill=white}]
\draw[csRule,line width=0.4pt] (19.50,-14.40) -- (19.50,-16.20);
\draw[csRule,line width=0.4pt] (19.50,-16.20) -- (19.50,-18.00);
\draw[csRule,line width=0.4pt] (27.30,-14.40) -- (27.30,-16.20);
\draw[csRule,line width=0.4pt] (27.30,-16.20) -- (27.30,-18.00);
\draw[csRule,line width=0.4pt] (31.20,-14.40) -- (31.20,-16.20);
\draw[csRule,line width=0.4pt] (31.20,-16.20) -- (31.20,-18.00);
\draw[csRule,line width=0.4pt] (23.40,-10.80) -- (23.40,-12.60);
\draw[csRule,line width=0.4pt] (15.60,-12.60) -- (31.20,-12.60);
\draw[csRule,line width=0.4pt] (15.60,-12.60) -- (15.60,-14.40);
\draw[csRule,line width=0.4pt] (19.50,-12.60) -- (19.50,-14.40);
\draw[csRule,line width=0.4pt] (23.40,-12.60) -- (23.40,-14.40);
\draw[csRule,line width=0.4pt] (27.30,-12.60) -- (27.30,-14.40);
\draw[csRule,line width=0.4pt] (31.20,-12.60) -- (31.20,-14.40);
\draw[csRule,line width=0.4pt] (23.40,-7.20) -- (23.40,-9.00);
\draw[csRule,line width=0.4pt] (23.40,-9.00) -- (23.40,-10.80);
\draw[csRule,line width=0.4pt] (23.40,-3.60) -- (23.40,-5.40);
\draw[csRule,line width=0.4pt] (11.70,-5.40) -- (35.10,-5.40);
\draw[csRule,line width=0.4pt] (11.70,-5.40) -- (11.70,-7.20);
\draw[csRule,line width=0.4pt] (23.40,-5.40) -- (23.40,-7.20);
\draw[csRule,line width=0.4pt] (35.10,-5.40) -- (35.10,-7.20);
\draw[csRule,line width=0.4pt] (42.90,-7.20) -- (42.90,-9.00);
\draw[csRule,line width=0.4pt] (42.90,-9.00) -- (42.90,-10.80);
\draw[csRule,line width=0.4pt] (42.90,-3.60) -- (42.90,-5.40);
\draw[csRule,line width=0.4pt] (42.90,-5.40) -- (42.90,-7.20);
\draw[csRule,line width=0.4pt] (23.40,0.00) -- (23.40,-1.80);
\draw[csRule,line width=0.4pt] (0.00,-1.80) -- (46.80,-1.80);
\draw[csRule,line width=0.4pt] (0.00,-1.80) -- (0.00,-3.60);
\draw[csRule,line width=0.4pt] (3.90,-1.80) -- (3.90,-3.60);
\draw[csRule,line width=0.4pt] (7.80,-1.80) -- (7.80,-3.60);
\draw[csRule,line width=0.4pt] (23.40,-1.80) -- (23.40,-3.60);
\draw[csRule,line width=0.4pt] (39.00,-1.80) -- (39.00,-3.60);
\draw[csRule,line width=0.4pt] (42.90,-1.80) -- (42.90,-3.60);
\draw[csRule,line width=0.4pt] (46.80,-1.80) -- (46.80,-3.60);
\node at (0.00,-3.60) {=};
\node[fill=csStore,rounded corners=0.8pt] at (3.90,-3.60) {=};
\node at (7.80,-3.60) {=};
\node[fill=csStore,rounded corners=0.8pt] at (11.70,-7.20) {=};
\node at (15.60,-14.40) {=};
\node at (19.50,-18.00) {cont};
\node at (19.50,-14.40) {if};
\node at (23.40,-14.40) {=};
\node at (27.30,-18.00) {=};
\node at (27.30,-14.40) {if};
\node at (31.20,-18.00) {=};
\node at (31.20,-14.40) {if};
\node at (23.40,-10.80) {for};
\node at (23.40,-7.20) {for};
\node[fill=csStore,rounded corners=0.8pt] at (35.10,-7.20) {=};
\node at (23.40,-3.60) {for};
\node at (39.00,-3.60) {=};
\node at (42.90,-10.80) {=};
\node at (42.90,-7.20) {for};
\node at (42.90,-3.60) {for};
\node at (46.80,-3.60) {ret};
\node at (23.40,0.00) {def};
\end{tikzpicture}\par}
\end{minipage}\hspace{3\csCW}%
\begin{minipage}[t]{\dimexpr 56\csCW+1pt\relax}
{\scriptsize \pminus{} (slower)\hfill\textbf{1255.0\,ms}}\par\vspace{1pt}
\begin{Verbatim}[commandchars=\\\{\},fontsize=\scriptsize,frame=lines,framerule=0.4pt,rulecolor=\color{csRule},framesep=2pt,
  listparameters={\setlength{\topsep}{0pt}\setlength{\partopsep}{0pt}\setlength{\parskip}{0pt}}]
MOD = 10**9 + 7
\csS{dp = [[[0] * 3 for _ in range(2)] for _ in range(n + 1)]}
dp\csI{[0]}[0][0] = 1
for \csI{i} in range(\csI{1, n + 1}):

    for a in range(2):
        for l in range(3):
            val = dp\csI{[i - 1]}[a][l]
            \csE{...}

\csE{...}
\end{Verbatim}
\par\vspace{4pt}{\centering
\begin{tikzpicture}[x=\csCW,y=\csCW,every node/.style={font=\scriptsize\ttfamily,inner sep=0.6pt,text height=1.6ex,text depth=0.3ex,fill=white}]
\draw[csRule,line width=0.4pt] (15.60,-14.40) -- (15.60,-16.20);
\draw[csRule,line width=0.4pt] (15.60,-16.20) -- (15.60,-18.00);
\draw[csRule,line width=0.4pt] (23.40,-14.40) -- (23.40,-16.20);
\draw[csRule,line width=0.4pt] (23.40,-16.20) -- (23.40,-18.00);
\draw[csRule,line width=0.4pt] (27.30,-14.40) -- (27.30,-16.20);
\draw[csRule,line width=0.4pt] (27.30,-16.20) -- (27.30,-18.00);
\draw[csRule,line width=0.4pt] (19.50,-10.80) -- (19.50,-12.60);
\draw[csRule,line width=0.4pt] (11.70,-12.60) -- (27.30,-12.60);
\draw[csRule,line width=0.4pt] (11.70,-12.60) -- (11.70,-14.40);
\draw[csRule,line width=0.4pt] (15.60,-12.60) -- (15.60,-14.40);
\draw[csRule,line width=0.4pt] (19.50,-12.60) -- (19.50,-14.40);
\draw[csRule,line width=0.4pt] (23.40,-12.60) -- (23.40,-14.40);
\draw[csRule,line width=0.4pt] (27.30,-12.60) -- (27.30,-14.40);
\draw[csRule,line width=0.4pt] (19.50,-7.20) -- (19.50,-9.00);
\draw[csRule,line width=0.4pt] (19.50,-9.00) -- (19.50,-10.80);
\draw[csRule,line width=0.4pt] (19.50,-3.60) -- (19.50,-5.40);
\draw[csRule,line width=0.4pt] (19.50,-5.40) -- (19.50,-7.20);
\draw[csRule,line width=0.4pt] (35.10,-7.20) -- (35.10,-9.00);
\draw[csRule,line width=0.4pt] (35.10,-9.00) -- (35.10,-10.80);
\draw[csRule,line width=0.4pt] (35.10,-3.60) -- (35.10,-5.40);
\draw[csRule,line width=0.4pt] (35.10,-5.40) -- (35.10,-7.20);
\draw[csRule,line width=0.4pt] (19.50,0.00) -- (19.50,-1.80);
\draw[csRule,line width=0.4pt] (0.00,-1.80) -- (39.00,-1.80);
\draw[csRule,line width=0.4pt] (0.00,-1.80) -- (0.00,-3.60);
\draw[csRule,line width=0.4pt] (3.90,-1.80) -- (3.90,-3.60);
\draw[csRule,line width=0.4pt] (7.80,-1.80) -- (7.80,-3.60);
\draw[csRule,line width=0.4pt] (19.50,-1.80) -- (19.50,-3.60);
\draw[csRule,line width=0.4pt] (31.20,-1.80) -- (31.20,-3.60);
\draw[csRule,line width=0.4pt] (35.10,-1.80) -- (35.10,-3.60);
\draw[csRule,line width=0.4pt] (39.00,-1.80) -- (39.00,-3.60);
\node at (0.00,-3.60) {=};
\node[fill=csStore,rounded corners=0.8pt] at (3.90,-3.60) {=};
\node at (7.80,-3.60) {=};
\node at (11.70,-14.40) {=};
\node at (15.60,-18.00) {cont};
\node at (15.60,-14.40) {if};
\node at (19.50,-14.40) {=};
\node at (23.40,-18.00) {=};
\node at (23.40,-14.40) {if};
\node at (27.30,-18.00) {=};
\node at (27.30,-14.40) {if};
\node at (19.50,-10.80) {for};
\node at (19.50,-7.20) {for};
\node at (19.50,-3.60) {for};
\node at (31.20,-3.60) {=};
\node at (35.10,-10.80) {=};
\node at (35.10,-7.20) {for};
\node at (35.10,-3.60) {for};
\node at (39.00,-3.60) {ret};
\node at (19.50,0.00) {def};
\end{tikzpicture}\par}
\end{minipage}\par\vspace{2pt}
{\scriptsize AST similarity of the pair 0.993\par\vspace{5pt}
$\Downarrow$\ two refinement rounds, \pminus{} the partner in both:\ 899.6\,ms $\rightarrow$ 561.8\,ms
$\rightarrow$ 100.3\,ms, 9.0$\times$ faster\par}\vspace{5pt}
\begin{minipage}{\dimexpr 108\csCW+4pt\relax}
{\scriptsize refined program (round 2, the $r_2$ pick)\hfill\textbf{100.3\,ms}}\par\vspace{1pt}
\vspace{2pt}\nointerlineskip
\begin{minipage}[c]{\dimexpr 59\csCW+2pt\relax}
\begin{Verbatim}[commandchars=\\\{\},fontsize=\scriptsize,frame=lines,framerule=0.4pt,rulecolor=\color{csRule},framesep=2pt,
  listparameters={\setlength{\topsep}{0pt}\setlength{\partopsep}{0pt}\setlength{\parskip}{0pt}}]
MOD = 10**9 + 7
\csS{a0l0, a0l1, a0l2 = 1, 0, 0}
\csS{a1l0, a1l1, a1l2 = 0, 0, 0}
for _ in range(n):
    na0l0 = (a0l0 + a0l1 + a0l2) % MOD
    na0l1 = a0l0
    na0l2 = a0l1
    na1l0 = (a1l0 + a1l1 + a1l2 + a0l0 + a0l1 + a0l2) % MOD
    na1l1 = a1l0
    na1l2 = a1l1
    \csS{a0l0, a0l1, a0l2 = na0l0, na0l1, na0l2}
    \csS{a1l0, a1l1, a1l2 = na1l0, na1l1, na1l2}
return (a0l0 + a0l1 + a0l2 + a1l0 + a1l1 + a1l2) % MOD
\end{Verbatim}
\end{minipage}\hspace{3\csCW}%
\begin{minipage}[c]{46\csCW}\centering
\begin{tikzpicture}[x=\csCW,y=\csCW,every node/.style={font=\scriptsize\ttfamily,inner sep=0.6pt,text height=1.6ex,text depth=0.3ex,fill=white}]
\draw[csRule,line width=0.4pt] (25.35,-3.60) -- (25.35,-5.40);
\draw[csRule,line width=0.4pt] (11.70,-5.40) -- (39.00,-5.40);
\draw[csRule,line width=0.4pt] (11.70,-5.40) -- (11.70,-7.20);
\draw[csRule,line width=0.4pt] (15.60,-5.40) -- (15.60,-7.20);
\draw[csRule,line width=0.4pt] (19.50,-5.40) -- (19.50,-7.20);
\draw[csRule,line width=0.4pt] (23.40,-5.40) -- (23.40,-7.20);
\draw[csRule,line width=0.4pt] (27.30,-5.40) -- (27.30,-7.20);
\draw[csRule,line width=0.4pt] (31.20,-5.40) -- (31.20,-7.20);
\draw[csRule,line width=0.4pt] (35.10,-5.40) -- (35.10,-7.20);
\draw[csRule,line width=0.4pt] (39.00,-5.40) -- (39.00,-7.20);
\draw[csRule,line width=0.4pt] (21.45,0.00) -- (21.45,-1.80);
\draw[csRule,line width=0.4pt] (0.00,-1.80) -- (42.90,-1.80);
\draw[csRule,line width=0.4pt] (0.00,-1.80) -- (0.00,-3.60);
\draw[csRule,line width=0.4pt] (3.90,-1.80) -- (3.90,-3.60);
\draw[csRule,line width=0.4pt] (7.80,-1.80) -- (7.80,-3.60);
\draw[csRule,line width=0.4pt] (25.35,-1.80) -- (25.35,-3.60);
\draw[csRule,line width=0.4pt] (42.90,-1.80) -- (42.90,-3.60);
\node at (0.00,-3.60) {=};
\node[fill=csStore,rounded corners=0.8pt] at (3.90,-3.60) {=};
\node[fill=csStore,rounded corners=0.8pt] at (7.80,-3.60) {=};
\node at (11.70,-7.20) {=};
\node at (15.60,-7.20) {=};
\node at (19.50,-7.20) {=};
\node at (23.40,-7.20) {=};
\node at (27.30,-7.20) {=};
\node at (31.20,-7.20) {=};
\node[fill=csStore,rounded corners=0.8pt] at (35.10,-7.20) {=};
\node[fill=csStore,rounded corners=0.8pt] at (39.00,-7.20) {=};
\node at (25.35,-3.60) {for};
\node at (42.90,-3.60) {ret};
\node at (21.45,0.00) {def};
\end{tikzpicture}
\end{minipage}
\end{minipage}\par
\end{minipage}%
\end{lrbox}%
\ifdim\wd\csPairBox>\linewidth\resizebox{\dimexpr\linewidth-1pt\relax}{!}{\usebox{\csPairBox}}%
\else\usebox{\csPairBox}\fi
\caption{Mercury \#552 (DeepSeek-V4.1 Flash): the round-1 pair and the final refined program, each with its
statement-level AST. Up to cosmetic edits, the pair differs only in how the table is stored (shaded) and the indexing
this forces (blue); the refined program keeps the six states in scalars. \texttt{...}~marks omitted lines. Full case in
Appendix~\ref{app:case552}.}
\label{fig:case552-pair}
\end{figure}

\subsection{Profiling Method Analysis}
\label{appendix:profiling}

\subsubsection{Compressed Profiling Signal}
\label{appendix:profiling-detail-compact}
We compact the profiler output to a short summary: the candidate function's own share of profiled CPU time -- cProfile attributes time per function, so this is reported at the function's definition line -- together with up to five of the hottest calls it makes, each accounting for at least 1\% of profiled time. Profiler frames and harness scaffolding are excluded so they cannot appear as spurious hotspots.

\subsubsection{Rationale for Compressed Profiling Signal}
\label{appendix:whycondensedprofiling}

Our goal in profiling is not to recover perfectly precise line-by-line execution costs. Rather, we need a lightweight and reliable signal that identifies where most of the runtime is spent, so that refinement can focus on the most promising optimization opportunities. This is why a condensed profiler signal is sufficient for our setting.

\begin{center}
\fbox{%
\parbox{0.92\linewidth}{%
\textbf{Amdahl's Law.}
If a fraction $p$ of total runtime is spent in one component, and that component is accelerated by a factor of $k$, then the overall speedup is
\[
S_{\mathrm{total}} = \frac{1}{(1-p) + \frac{p}{k}}.
\]
}
}
\end{center}

Amdahl's law shows that the end-to-end benefit of an optimization depends primarily on how much of the total runtime is covered by the optimized component. In particular, larger values of $p$ yield larger global gains. This formalizes the intuition to \emph{optimize the hottest code first}: improving a small bottleneck cannot produce a large overall speedup, no matter how aggressively it is optimized.

This observation is especially relevant for our use case. Since the refinement loop only needs to know which functions or lines dominate execution, approximate hotspot attribution is often enough. 

We can also view each refinement step as a greedy optimization move. Let $J$ denote the total execution cost. Choosing the hotspot with the largest estimated contribution to $J$ is analogous to taking a local step in the direction of steepest decrease. Even when profiling provides only relative percentages rather than exact measurements, these percentages still supply a useful ordering over candidate bottlenecks.



\subsection{Prompts}

\subsubsection{\effipair Generation}
\lstset{
  language=Python,
  basicstyle=\ttfamily\small,
  keywordstyle=\color{blue},
  commentstyle=\color{gray},
  stringstyle=\color{red},
  showstringspaces=false,
  frame=single,
  breaklines=true
}

  \textbf{\effipair Generation (ENAMEL)}
  \begin{lstlisting}
  Complete the following function.

  {prompt_stub}

  Requirements:
  - Keep the required function signature (top-level, no class).
  - Return only a Python code block.
  \end{lstlisting}

  \textbf{\effipair Generation (EvalPerf)}
  \begin{lstlisting}
  {prompt_text}

  Requirements:
  - Implement the top-level function named '{entry_point}' (no class wrapper).
  - Return only a Python code block.
  \end{lstlisting}

  \textbf{\effipair Generation (Mercury)}
  \begin{lstlisting}
  Task:
  {pretty_problem_content}

  Code Stub:
  {stub}

  Requirements:
  - Provide class Solution with the required method '{method_name}'.
  - Return only a Python code block.
  \end{lstlisting}

\subsubsection{\effipair Efficiency}

  \begin{lstlisting}
  Improve the following Python code with correctness as the top priority and efficiency as secondary.

  Guidance:
  {guidance}

  {task_block}
  Candidate A:
  {code_a}

  [optional] Profile A:
  {profile_a}

  [optional] Candidate B:

  {code_b}

  [optional] Profile B:
  {profile_b}

  Requirements:
  {requirements}
  \end{lstlisting}

\subsubsection{EffiLearner Generation}

\textbf{HumanEval.}
  \begin{lstlisting}
  Please complete Python code based on the task description. # Task description:
  {prompt}
  #Solution:
  \end{lstlisting}

\textbf{MBPP.}
  \begin{lstlisting}
  Please complete Python code based on the task description and test cases. # Task description:
  {prompt}
  {tests}
  #Solution:
  \end{lstlisting}

\subsubsection{EffiLearner Efficiency}
  \begin{lstlisting}
  Optimize the efficiency of the following Python code based on the task, test case, and overhead analysis provided. Ensure the optimized code can pass the given test
  case.

  Task Description:
  {task_description}

  Test Case:
  {test_case}

  Original Code:

  {completion}

  Overhead Analysis:
  {overhead_prompt}

  Optimization Rules:

  - Encapsulate the optimized code within a Python code block (i.e., python\n[Your Code Here]\n).
  - Do not include the test case within the code block.
  - Focus solely on code optimization; test cases are already provided.
  - Ensure the provided test case passes with your optimized solution.
    \end{lstlisting}

\subsection{Case study: Mercury task \#552}
\label{app:case552}

This appendix follows one Mercury task, \#552 (\texttt{student-attendance-record-ii}) as a representative example of how \effipair turns differences between structurally similar programs into optimization guidance. The task uses the entry point \texttt{checkRecord} and is labeled LeetCode Hard and asks for the number of valid attendance sequences of length n with fewer than two absences and no three consecutive late days. Figure~\ref{fig:case552-pair} summarizes the selected pair and the final refined implementation.

Each draw runs on Mercury's offline checker over the 8 test cases, and three clocks are recorded: the wall time of that correctness run, the workload time (in-process, candidate calls only, mean of 3 runs) and the profile time (the elapsed time under cProfile). In this run the pair is ordered by the wall time, the prompt reports the profile time and the committed pick uses the workload time. The naive draw enumerates all $3^n$ records, reaches the 30\,s limit and is excluded from pairing (Table~\ref{tab:c552-gate}). The other two draws are the same dynamic program over (absences, current late streak): the \emph{medium} draw keeps the full $(n{+}1)\times2\times3$ table and the \emph{fast} draw keeps two rolling $2\times3$ rows.

\begin{table}[!ht]
\centering\small
\caption{Gate and clocks of the three draws (iteration 0).}
\label{tab:c552-gate}
\begin{tabular}{llllrr}
\toprule
draw & style & gate & wall (s) & workload (ms) & profile (ms) \\
\midrule
\texttt{1d7bfd0f} & medium & ok & 1.156 & 1085.6 & 1255.0 \\
\texttt{eb72593a} & fast & ok & 0.797 & 728.4 & 899.6 \\
\texttt{09cc3033} & naive & timeout & 30.031 & -- & -- \\
\bottomrule
\end{tabular}
\end{table}

\subsubsection{Refinement}
\paragraph{Round One.} \effipair selects the fastest correct candidate \(p^+\) and the slowest correct candidate \(p^-\) within the similarity window. Here, \(p^+=\texttt{eb72593a}\) and \(p^-=\texttt{1d7bfd0f}\), with AST similarity \(0.993\). As shown in Figure \ref{fig:case552-pair}, the programs implement the same six-state dynamic program and differ primarily in data layout: \(p^+\) retains only two rolling \(2\times3\) tables, whereas \(p^-\) stores the full \((n+1)\times2\times3\) table. Their profile times are 899.6 ms and 1255.0 ms, respectively, providing a localized performance contrast between otherwise highly similar implementations. Conditioned on this pair and its execution feedback, the first refinement preserves the rolling-table representation but removes the nested loops over the six states, replacing them with explicit state updates. The resulting candidate, \(\texttt{49b12e17}\), passes the correctness gate and reduces profile time from 899.6 ms to 561.8 ms; its workload time decreases from 728.4 ms to 523.4 ms.

\paragraph{Round Two.} The round-1 output becomes the new reference candidate \(p^+\). The same slower candidate \(\texttt{1d7bfd0f}\) remains the selected partner \(p^-\), with AST similarity \(0.969\). The second refinement further simplifies the representation by replacing the two \(2\times3\) lists with six scalar state variables, eliminating per-iteration list allocation and indexing. This candidate, \(\texttt{a4522979}\), also passes the correctness gate and reduces profile time from 561.8 ms to 100.3 ms, with workload time decreasing from 523.4 ms to 88.5 ms.

Across the two rounds, the selected implementation therefore progresses from a rolling-table dynamic program to explicit state transitions and finally to a scalar-state formulation. Profile time decreases from 899.6 ms to 100.3 ms while correctness is preserved. The example illustrates the intended role of DEF: the paired program supplies a nearby implementation contrast, while successive refinement can exploit that contrast to discover additional optimizations beyond simply copying the faster candidate.

\begin{lstlisting}[style=csPy,basicstyle=\ttfamily\tiny,caption={Round-1 output \texttt{49b12e17} (the program inside the model's code block).},label=lst:c552-49b12e17]
class Solution(object):
    def checkRecord(self, n):
        MOD = 10**9 + 7

        # dp[a][l]
        # a = total absences so far: 0 or 1
        # l = current consecutive late streak length: 0, 1, or 2
        dp = [[0] * 3 for _ in range(2)]
        dp[0][0] = 1

        for _ in range(n):
            ndp = [[0] * 3 for _ in range(2)]

            # Unroll the 6 states to reduce loop overhead while preserving correctness.
            v = dp[0][0]
            if v:
                ndp[0][0] = (ndp[0][0] + v) % MOD       # P
                ndp[1][0] = (ndp[1][0] + v) % MOD       # A
                ndp[0][1] = (ndp[0][1] + v) % MOD       # L

            v = dp[0][1]
            if v:
                ndp[0][0] = (ndp[0][0] + v) % MOD       # P
                ndp[1][0] = (ndp[1][0] + v) % MOD       # A
                ndp[0][2] = (ndp[0][2] + v) % MOD       # L

            v = dp[0][2]
            if v:
                ndp[0][0] = (ndp[0][0] + v) % MOD       # P
                ndp[1][0] = (ndp[1][0] + v) % MOD       # A

            v = dp[1][0]
            if v:
                ndp[1][0] = (ndp[1][0] + v) % MOD       # P
                ndp[1][1] = (ndp[1][1] + v) % MOD       # L

            v = dp[1][1]
            if v:
                ndp[1][0] = (ndp[1][0] + v) % MOD       # P
                ndp[1][2] = (ndp[1][2] + v) % MOD       # L

            v = dp[1][2]
            if v:
                ndp[1][0] = (ndp[1][0] + v) % MOD       # P

            dp = ndp

        return (dp[0][0] + dp[0][1] + dp[0][2] + dp[1][0] + dp[1][1] + dp[1][2]) % MOD
\end{lstlisting}

\subsubsection{Pool, committed picks and official score}
Table~\ref{tab:c552-pool} lists the whole pool. The committed pick after round $r$ is the fastest candidate the pipeline
marked correct, on the workload clock, among iterations $\le r$. Each pick is scored with Mercury's procedure (median of
3 repeats, \texttt{PYTHONHASHSEED=0}, the 8 test cases), and Beyond@1 $= (t_{\max} - t)/(t_{\max} - t_{\min})$,
clipped to $[0,1]$, against the 16 reference runtimes (Table~\ref{tab:c552-picks}). The results tables report
$r_0$ and $r_2$; the $r_1$ pick is included here for the round-by-round view. These three scores come from a separate
scoring session, so their absolute runtimes differ from the table's by that session's machine offset; the ordering is
the point. The $r_2$ program is faster than 14 of the 16 reference solutions.

\begin{table}[!ht]
\centering\small
\caption{Every entry of the pool with its clocks.}
\label{tab:c552-pool}
\setlength{\tabcolsep}{4pt}
\begin{tabular}{lrlllrrr}
\toprule
run & iter. & source & Cand.\ B & gate & wall (s) & workload (ms) & profile (ms) \\
\midrule
\texttt{09cc3033} & 0 & naive draw & -- & timeout & 30.031 & -- & -- \\
\texttt{1d7bfd0f} & 0 & medium draw & -- & ok & 1.156 & 1085.6 & 1255.0 \\
\texttt{eb72593a} & 0 & fast draw & -- & ok & 0.797 & 728.4 & 899.6 \\
\texttt{49b12e17} & 1 & refines \texttt{eb72593a} & \texttt{1d7bfd0f} & ok & 0.593 & 523.4 & 561.8 \\
\texttt{a4522979} & 2 & refines \texttt{49b12e17} & \texttt{1d7bfd0f} & ok & 0.157 & 88.5 & 100.3 \\
\bottomrule
\end{tabular}
\end{table}

\begin{table}[!ht]
\centering\small
\caption{Committed picks and their official Mercury scores.}
\label{tab:c552-picks}
\begin{tabular}{lllrr}
\toprule
pick & run & passes & runtime (s, median of 3) & Beyond@1 \\
\midrule
$r_0$ & \texttt{eb72593a} & yes & 0.3756 & 54.0 \\
$r_1$ & \texttt{49b12e17} & yes & 0.2760 & 66.2 \\
$r_2$ & \texttt{a4522979} & yes & 0.0508 & 94.0 \\
\bottomrule
\end{tabular}
\end{table}

\subsubsection{Other repetitions and cost}
\label{app:case552-reps}
Repetitions 1 and 2 are independent generations (seeds 136--138 and 146--148). In each repetition the naive draw
enumerates all $3^n$ records and fails the gate (it reaches the 30\,s limit in repetition~3 and fails in under
0.1\,s in repetitions 1 and~2), and the other two draws are the same dynamic program. Two rounds take the best program
from 0.90--1.26\,s to 0.10--0.20\,s on the profile clock (Table~\ref{tab:c552-reps}). In repetition~2 the pair again
contrasts rolling rows with the full table; in repetition~1 both draws keep the full table. Table~\ref{tab:c552-tokens}
gives the token cost of repetition~3.

\begin{table}[!ht]
\centering\small
\caption{Profile time (ms) of the best correct program after each stage in the three repetitions, the partner
(Candidate B) of both rounds, and the wall time (s) of the naive draw's gate run. $^{\ast}$The repetition shown above.}
\label{tab:c552-reps}
\setlength{\tabcolsep}{4pt}
\begin{tabular}{llllll}
\toprule
rep. & best draw ($r_0$) & after round 1 & after round 2 & partner & naive (s) \\
\midrule
1 & \texttt{1cefd2c5} (medium) 1258.7 & \texttt{bd58c82f} 999.2 & \texttt{6b512ed5} 162.3 & \texttt{aa9818e2} & 0.071 \\
2 & \texttt{e543525f} (fast) 957.3 & \texttt{31af93ac} 313.2 & \texttt{c8225cda} 195.4 & \texttt{0e89ba0d} & 0.064 \\
3$^{\ast}$ & \texttt{eb72593a} (fast) 899.6 & \texttt{49b12e17} 561.8 & \texttt{a4522979} 100.3 & \texttt{1d7bfd0f} & 30.031 \\
\bottomrule
\end{tabular}
\end{table}

\begin{table}[!ht]
\centering\small
\caption{Tokens for this task in repetition 3.}
\label{tab:c552-tokens}
\begin{tabular}{lrr}
\toprule
stage & prompt & completion \\
\midrule
generation, three draws ($r_0$) & 1139 & 799 \\
refinement round 1 & 1276 & 606 \\
refinement round 2 & 1528 & 414 \\
\midrule
total ($r_2$) & 3943 & 1819 \\
\bottomrule
\end{tabular}
\end{table}

\subsection{Repository-Level Task Exclusion Details}
\label{app:swe-perf-exclusions}

We evaluate on 125 of SWE-Perf's 140 instances. Before running any arm we replayed the benchmark's gold patch three times per instance under the unmodified harness. Ten instances did not survive that check. On eight (\texttt{seaborn-2224}, \texttt{seaborn-2313}, \texttt{scikit-learn-13124}, \texttt{scikit-learn-12446}, \texttt{seaborn-2458}, \texttt{requests-1891}, \texttt{astropy-13734} and \texttt{scikit-learn-12246}) every repetition ran cleanly, but the expert patch measured a gain of $0.0\%$ each time against the $2.9$--$20.8\%$ the benchmark records, leaving no reference to compare an arm against. Two more, \texttt{sphinx-8537} and \texttt{sphinx-6422}, produced no timing at all in any repetition. The remaining five are limits of our patch builder rather than of the benchmark, and all five reproduce normally. \texttt{sympy-27347}, \texttt{scikit-learn-12859} and \texttt{astropy-15228} carry an empty \texttt{patch\_functions} field, so the oracle setting specifies no target functions to give the model. In \texttt{sympy-27666} and \texttt{scikit-learn-12567} every named target resolves to several definitions in its file, which our builder cannot splice unambiguously. No result in this paper is credited with any of the fifteen.

\end{document}

%% file: math_commands.tex
\usepackage{amsmath,amsfonts,bm}

\def\eqref#1{equation~\ref{#1}}

\def\1{\bm{1}}

\DeclareMathAlphabet{\mathsfit}{\encodingdefault}{\sfdefault}{m}{sl}
\SetMathAlphabet{\mathsfit}{bold}{\encodingdefault}{\sfdefault}{bx}{n}

